\documentclass[11pt]{article}
\usepackage{graphicx}

\begin{document}

\title{\textbf{\textsf{Interacting Modified Variable Chaplygin Gas in a Non-flat Universe}}}
\author{ Mubasher Jamil\footnote{Corresponding author: mjamil@camp.edu.pk}\ \ and Muneer Ahmad Rashid\footnote{muneerrshd@yahoo.com}
\\ \\
%EndAName
\textit{\small Center for Advanced Mathematics and Physics}\\
\textit{\small National University of Sciences and Technology}\\
\textit{\small Peshawar Road, Rawalpindi - 46000, Pakistan} \\
%EndAName
} \maketitle
\begin{abstract}
A unified model of dark energy and matter is presented using the
modified variable Chaplygin gas for interacting dark energy in a
non-flat universe. The two entities interact with each other
non-gravitationally which involves a coupling constant. Due to
dynamic interaction, the variation in this constant arises that
henceforth changes the equations of state of these quantities. We
have derived the effective equations of state corresponding to
matter and dark energy in this interacting model. Moreover, the case
of phantom energy is deduced by putting constraints on the
parameters involved.
\end{abstract}


\section{Introduction}
Hosts of results coming from the observations of WMAP
\cite{sper,wang} have convincingly shown the validity of standard
Big Bang model with cosmological constant $\Lambda$, so called
$\Lambda$CDM model. Surprisingly the energy density corresponding to
$\Lambda$ is two third of the critical density or
$\Omega_\Lambda\approx0.7$, the remaining part is due to baryonic
and non-baryonic dark matter. The genesis of this energy, commonly
called the dark energy, and its subsequent evolution poses serious
problems to investigate. The dark energy is commonly represented by
a barotropic equation of state (EoS) $p=-\rho$ (or $\omega=-1$).
Numerous models have been proposed in the literature to explain the
cosmic accelerated expansion which include those based on
homogeneous and time dependent scalar field termed as quintessence
\cite{wang1}, phantom energy having EoS $\omega<-1$ \cite{cald},
Chaplygin gas (CG) with the EoS $p=-X/\rho$ \cite{kamen},
dissipative or viscous cosmology \cite{ren} and modified gravity
models including Dvali-Gabadadze-Porrati (DGP) model \cite{dvali}
and higher order $f(R)$ gravity theories \cite{nojiri}.

Among all of these alternative theories, the Chaplygin gas
effectively explains the evolution of the universe right from the
earlier matter dominated era (decelerating phase) to the late dark
energy dominated era (accelerating phase). This CG was first
introduced in the aerodynamical context \cite{chap} and has been
largely investigated in recent years due to its adequate efficiency
in interpolating the astrophysical data.

The energy conservation equation $\dot{\rho}+3H(\rho+p)=0$ yields
\begin{equation}
\rho=\sqrt{X+\frac{Y}{a^6}},
\end{equation}
where $X$ is a positive constant and $Y$ is the constant of
integration. Above, $p$ and $\rho$ are the pressure and energy
densities of the CG. Thus small values of $a(t)$ mimic dust phase
while large values represent dark energy phase of the universe. The
CG model favors a spatially flat universe with 95.4 \% confidence
level which agrees with the observational data of Sloan Digital Sky
Survey (SDSS) and Supernova Legacy Survey (SNLS) \cite{xun}. It also
possesses a property of giving accelerated expansion even if it gets
coupled with other scalar fields like quintessence\ or dissipative\
matter fields \cite{Chimento1}. Due to its this effectiveness,
various generalizations of CG have been proposed in the literature
(see e.g.
\cite{bento,cruz,bena,seta1,seta2,seta3,hong,seta4,guo,debna}). All
these models based on the inhomogeneous Chaplygin gas offer unified
picture of dark matter and dark energy (sometimes called dark
matter-dark energy unification) \cite{bilic}. But at the same time
CG has a drawback as it produces oscillations or exponential blow up
of dark matter power spectrum, which is inconsistent with the
observations \cite {sand}. Similar results have been obtained in
later generalizations of CG \cite{Carturan}.

One of the long standing problems in cosmology for the past decade
is the so-called ``cosmic coincidence problem'' which naively asks:
If the universe evolves from the earlier quintessence ($\omega>-1$)
to the later phantom regime ($\omega<-1$), then why $\omega=-1$
crossing is occurring at the present time. The astrophysical data
seems to favor the evolving dark energy in which the EoS parameter
$\omega$ changes at different cosmic epochs \cite{sad}. This problem
can be rephrased alternatively as: In any cosmological model
explaining dark energy, the ratio of matter to dark energy density
$r_m$ is expected to decrease with the expansion of the universe.
However the observations suggest that this ratio is of the order
unity. It may implies the transfer of energy between these two
entities to keep a delicate balance at the current time. The models
in which this interaction is investigated are called the
`interacting dark energy models'. The interaction involves a
coupling constant that determines the strength of this interaction.

Earlier, it had been suggested that interacting models of CG with
dark matter can effectively ameliorate the coincidence problem which
is not possible in pure CG models \cite{zhang,wu,campo}. Moreover,
this problem has also been investigated in the context of
dissipative CG and successfully explains the cosmic conundrum
problem \cite{Chimento1}. The interacting model effectively explains
the phantom divide $\omega=-1$ i.e. the transition from $\omega>-1$
to $\omega<-1$ \cite{sad,seta5,seta6}. It also rules out the
possibility of the exotic phenomenon termed `Big Rip' which is the
infinite expansion of the universe in a finite time \cite{wu1}. It
also yields stable stationary attractor solution of the
Friedmann-Robertson-Walker (FRW) equations at late times. So if dark
energy evolves then it would be worthwhile to find its effective
EoS. This motivates us to investigate the more generalized form of
interacting CG and to find the effective EoS of dark energy and
matter in the respective model. The formalism adopted here is from
Ref. \cite{seta4}.

\section{Interacting modified variable Chaplygin gas}

We assume the background spacetime to be spatially homogeneous and
isotropic represented by the FRW metric
\begin{equation}
ds^2=-dt^2+a^2(t)\left[\frac{dr^2}{1-kr^2}+r^2(d\theta^2+\sin^2\theta
d\phi^2)\right].
\end{equation}
Here $a(t)$ is the scale factor and $k=-1,0,+1$ represent spatially
hyperbolic, flat and closed universes respectively. The
corresponding Einstein field equation is
\begin{equation}
H^2+\frac{k}{a^2}=\frac{1}{3M_p^2}(\rho_\Lambda+\rho_m),
\end{equation}
where $M^2_p=(8\pi G)^{-1}$ is the modified Planck mass. Moreover,
the energy conservation for our dynamical system is given by
\begin{equation}
\dot{\rho}_{\Lambda}+\dot{\rho}_m+3H[(1+\omega_\Lambda)\rho_\Lambda+\rho_m]=0,
\end{equation}
where we have taken the EoS for the dark energy
$p_\Lambda=\omega_\Lambda \rho_\Lambda$ and for the matter part
$p_m=0$ (or $\omega_m=0$).

To derive the general form of the interaction between CG and matter,
we start by considering the action of the scalar-tensor theory of
gravitation given by
\begin{equation}
S_{ST}=\int d^4x\sqrt{-g}\left[
\frac{\Re}{2}-\frac{1}{2}\partial_\mu\phi\partial^\mu\phi+\frac{1}{\Theta^2(\phi)}L_{m}(\xi,\partial\xi,\Theta^{-1}g_{\mu\nu})
\right],
\end{equation}
where $\Re$ is the Ricci scalar, $\phi$ is the spatially homogeneous
scalar field, $L_{m}$ represents the matter Lagrangian, $\xi$ is the
collective degrees of freedom of matter and $\Theta^{-2}$ is the
coupling function between matter and the scalar field. Using the
action $S_{ST}$, the interaction term $Q$ can be written as
\cite{zhang,curb}
\begin{equation}
Q=\rho_{m}H\frac{d\ln \Theta'}{d\ln a},
\end{equation}
where $\Theta'\equiv \Theta(a)^{(3\omega_{m}-1)/2}$. Letting
\begin{equation}
\Theta'(a)=\Theta_o e^{3\int
b^2\left(\frac{\rho_{m}+\rho_{\Lambda}}{\rho_{m}}\right)d\ln a}.
\end{equation}
Substitution of Eq. (7) in (6) yields
\begin{equation}
Q =3Hb^2(\rho _{\Lambda}+\rho _{m}),
\end{equation}
which is the general form of the interaction between CG and matter.
Here $b^2$ is the dimensionless coupling constant (also called
`transfer strength') for the interaction. The positive $b^2$ is
responsible for the transition from CG to matter and vice versa for
negative $b^2$. Sometimes this constant is taken in the range $[0,
1]$ (see Ref. \cite{zhang}). Note that if $b^2=0$ then it represents
the non-interacting FRW model while $b^2=1$ yields complete transfer
of energy from CG to matter. Recently, it is reported that this
interaction is observed in the Abell cluster A586 showing a
transition of dark energy into dark matter and vice versa
\cite{berto1}. Therefore the theoretical interacting models are
phenomenologically consistent with the observations.

Let us assume an interaction $Q=\Gamma\rho_\Lambda$ between CG and
dark matter, where $\Gamma$ is decay rate. Then Eq. (4) can be
divided into two parts corresponding to $\rho_\Lambda$ and $\rho_m$
as
\begin{eqnarray}
\dot{\rho}_{\Lambda}+3H(1+\omega_\Lambda)\rho_\Lambda&=&-Q,\\
\dot{\rho}_m+3H\rho_m&=&Q,
\end{eqnarray}
respectively. Eqs. (9) and (10) show that the energy conservation
for dark energy and matter would not hold independently if there is
an interaction between them but hold for the total interacting
system as manifested through Eq. (4). Further, we define the density
ratio $r_m$ as $r_m\equiv\frac{\rho_m}{\rho_\Lambda}$. To study how
this density ratio evolves with time, we differentiate $r_m$ with
respect to $t$ and obtain
\begin{equation}
\dot{r}_m=\frac{dr_m}{dt}=\frac{\rho_{m}}{\rho_{\Lambda}}\left[\frac{\dot{\rho}_{m}}{\rho_{m}}-\frac{\dot{\rho}_{\Lambda}}{\rho_{\Lambda}}\right],
\end{equation}
Using Eqs. (9) and (10) in (11), we get
\begin{equation}
\dot{r}_m=3Hr_m\left[\omega_\Lambda+\frac{1+r_m}{r_m}\frac{\Gamma}{3H}\right].
\end{equation}
Defining the effective EoS for dark energy and matter as \cite{kim}
\begin{equation}
\omega_\Lambda^{eff}=\omega_\Lambda+\frac{\Gamma}{3H}, \ \
\omega_m^{eff}=\frac{-1}{r_m}\frac{\Gamma}{3H},
\end{equation}
which involve the contribution from the interaction between matter
and dark energy. Using Eq. (13) in Eqs. (9) and (10), we get
\begin{eqnarray}
\dot{\rho}_{\Lambda}+3H(1+\omega_\Lambda^{eff})\rho_\Lambda&=&0,\\
\dot{\rho}_m+3H(1+\omega_m^{eff})\rho_m&=&0.
\end{eqnarray}
From the standard FRW model, the density parameters corresponding to
matter, dark energy and curvature are defined as
\begin{eqnarray}
\Omega_m&=&\frac{\rho_m}{\rho_{cr}}=\frac{\rho_m}{3H^2M^2_p},\\
\Omega_\Lambda&=&\frac{\rho_\Lambda}{\rho_{cr}}=\frac{\rho_\Lambda}{3H^2M^2_p},\\
\Omega_k&=&\frac{k}{a^2H^2}.
\end{eqnarray}
The parameter $\Omega_k$ represents the contribution in the total
energy density from the spatial curvature and is constrained as
$-0.0175<\Omega_k<0.0085$ with $95\%$ confidence level by current
observations \cite{water}. It is shown that a non-zero positive
curvature parameter $k$ allows for a bounce, thereby preventing the
cosmic singularities without violating the null energy condition
$\rho+p\geq0$ \cite{paris}.

The density parameters in Eqs. (16), (17) and (18) are related as
\begin{equation}
\Omega_m+\Omega_\Lambda=1+\Omega_k.
\end{equation}
Using definition of $r_m$ and Eq. (19), we can write
\begin{equation}
r_m=\frac{1+\Omega_k-\Omega_\Lambda}{\Omega_\Lambda}.
\end{equation}
In our generalized model, we choose the modified variable Chaplygin
gas (MVG) given by
\begin{equation}
p_{\Lambda}=A\rho_{\Lambda}-\frac{B(a)}{\rho_{\Lambda}^\alpha},\ \
B(a)=B_{o}a^{-n}.
\end{equation}
Here $A$, $B_{o}$ and $n$ are positive constants and
$0\leq\alpha\leq1$. Notice that for $B=0$, Eq. (21) represents the
barotropic EoS (or barotropic fluid). In general, the barotropic EoS
$p=A\rho$ represents various types of mediums. For instance, $A=-1$
gives cosmological constant or de Sitter vacuum; for $A=-2/3$, it
corresponds to domain walls; if $A=-1/3$ it gives cosmic strings;
while $A=0$ corresponds to dust or matter; $A=1/3$ yields the EoS
for relativistic gas like photon radiation; $A=2/3$ gives the
perfect gas and $A=1$ represents the ultra-stiff matter \cite{dymn}.
If $B(a)=B_{o}$ in Eq. (21), it gives EoS of modified CG. Again if
$A=0$ and $\alpha=1$, it represents the usual CG. Recently it is
deduced using the latest supernovae data that models with $\alpha>1$
are also possible \cite{berto}. Notice that for $A=0$, Eq. (21)
yields a fluid with negative pressure. Such a fluid is generally
characterized in the quintessence regime.

The modified form of the CG is also phenomenologically motivated and
can explain the flat rotational curves of galaxies \cite{tekola}.
The galactic rotational velocity $V_c$ is related with the MVG
parameter $A$ as $V_c=\sqrt{2A}$ and the density parameter $\rho$ is
related to the radial size of the galaxy as $\rho=\frac{A}{2\pi G
r^2}$. At high densities, the first term in MVG dominates and gives
the flat rotational curve that is consistent with with observations
for galaxies. The parameter $A$ varies from galaxy to galaxy due to
variations in $V_c$.

The density evolution of MVG is given by
\begin{equation}
\rho_\Lambda=(\Delta a^{-n}+Ca^{-s})^\frac{1}{1+\alpha},
\end{equation}
where $s\equiv3(1+\alpha)(1+A)$,
$\Delta\equiv\frac{3(1+\alpha)B_{o}}{s-n}$ and $C$ is the constant
of integration. Differentiating Eq. (22) we get
\begin{equation}
\dot{\rho}_\Lambda=\frac{-H}{1+\alpha}[(\Delta
a^{-n}+Ca^{-s})^\frac{-\alpha}{1+\alpha} (\Delta n a^{-n}+Csa^{-s})]
\end{equation}
Further, choosing the decay rate to be \cite{wang2}
\begin{equation}
\Gamma=3b^2H(1+r_m).
\end{equation}
Substituting Eq. (24) in Eq. (9) and using
$Q=3b^2H(1+r_m)\rho_\Lambda$, we obtain
\begin{equation}
\omega_\Lambda=\frac{1}{3(1+\alpha)}[(\Delta a^{-n}+Ca^{-s})^{-1}
(\Delta na^{-n}+Csa^{-s})] -b^2\frac{1+\Omega_k}{\Omega_\Lambda}-1.
\end{equation}
Using Eq. (20) and (24), we have
\begin{equation}
\Gamma=3b^2H\frac{1+\Omega_k}{\Omega_\Lambda}.
\end{equation}
Thus using Eqs. (14), (25) and (26), we arrive at the effective EoS
for dark energy as
\begin{equation}
\omega_\Lambda^{eff}=\frac{1}{3(1+\alpha)}[(\Delta
a^{-n}+Ca^{-s})^{-1} (\Delta na^{-n}+Csa^{-s})]-1.
\end{equation}
Notice that the cosmological models with phantom energy arise when
$\omega_\Lambda^{eff}<-1$ which is possible if the quantity in the
square brackets in Eq. (27) is less then zero. Therefore, we have
two possible cases:

Case (1). If $(\Delta a^{-n}+Ca^{-s})>0$ and $(\Delta
na^{-n}+Csa^{-s})<0$, then we have a constraint on $a(t)$
\begin{equation}
\frac{-C}{\Delta}<a^{s-n}<\frac{-Cs}{\Delta n}.
\end{equation}
This is possible when $\frac{-C}{\Delta}>0$ and $s>n$.

Case (2). If $(\Delta a^{-n}+Ca^{-s})<0$ and $(\Delta
na^{-n}+Csa^{-s})>0$, then we have
\begin{equation}
\frac{-Cs}{\Delta n}<a^{s-n}<\frac{-C}{\Delta},
\end{equation}
here $a(t)$ will be restricted when $\frac{-C}{\Delta}>0$ and $s<n$.

Thus the phantom dark energy case is possible if scale factor $a(t)$
is constrained as given in Eqs. (28) and (29). Also, Eq. (27)
represents usual cosmological constant if the quantity in the square
brackets vanishes identically.

Also we note that the quantity in Eq. (22) must be positive so that
$a(t)>(\frac{-C}{\Delta})^\frac{1}{s-n}$. Therefore, the minimum
value of $a(t)$ lies at
\begin{equation}
a_{min}=\left(\frac{-C}{\Delta}\right)^\frac{1}{s-n}.
\end{equation}
Therefore the universe can experience a bounce before reaching the
singularity when $a>a_{min}$ which corresponds to the bouncing
universe model. This bounce is possible due to our assumption of
positive curvature (see Ref. \cite{nove} for a review on bouncing
cosmologies). Moreover, the effective EoS of matter is determined
from Eqs. (13), (20) and (24) as
\begin{equation}
\omega^{eff}_m=-b^2\left(1+\frac{\Omega_\Lambda}{\Omega_m}\right).
\end{equation}
Using $\Omega_\Lambda=0.7$ and $\Omega_m=0.3$, we find from Eq. (31)
\begin{equation}
\omega^{eff}_m=-3.33b^2.
\end{equation}
Note that $\omega^{eff}_m$ depends only on the coupling constant. If
there is no interaction then $\omega^{eff}_m=0$. In particular if
$0\leq b^2\leq 1$, it implies $-3.33\leq\omega^{eff}_m\leq0$.
Therefore in the interacting model, matter can also behave as a
fluid with negative pressure which can cause the cosmic
acceleration.

\section{Conclusion}
In summary, we have determined the effective equations of state for
matter and dark energy in the generalized model of interacting
Chaplygin gas. In particular for dark energy, the modified variable
Chaplygin gas equation of state is used. Since dark energy interacts
with matter then due to energy exchange, the EoS of the both the
quantities will be modified. The effective EoS of Chaplygin gas
$\omega_\Lambda^{eff}$ also describes the EoS of phantom energy if
the scale factor is constrained by either Eq. (28) or (29). As
discussed, the interaction is observed in the near galactic cluster
Abell A586, we may hope to determine the effective EoS of both
matter and dark energy as well through future astrophysical
observations that will finally put further constraints on the
parameters in the model. Also it is shown that a universe filled
with the Chaplygin gas can bounce from the singularity if $a(t)$ is
greater then $a_{min}$, where the later describes the minimum value
of scale factor at the bounce. Finally, it is shown that matter can
also generate negative pressure along with dark energy to cause
cosmic expansion.

\subsubsection*{Acknowledgment}
We would like to thank anonymous referees for giving useful comments
to improve this work.

\end{document}